\documentstyle[twoside,fleqn,espcrc2]{article}
\input epsf

\title{Finite volume effects and quenched chiral logarithms}

\author{Robert D. Mawhinney
  \address{Department of Physics, Columbia
  University, New York, NY, 10027, U.S.A}
  \thanks{This work was done in collaboration with Shailesh
  Chandrasekharan, Dong Chen, Norman H. Christ, and Weonjong
  Lee.  It was supported in part by the US Department of
  Energy.}}
       
\begin{document}

\def\thepage{CU--TP--738}
\thispagestyle{myheadings}

\begin{abstract}
We have measured the valence pion mass and the valence chiral
condensate on lattice configurations generated with and without
dynamical fermions.  We find that our data and that of others is well
represented by a linear relationship between $m_{\pi}^2$ and the
valence quark mass, with a non-zero intercept.  For our data, we relate
the intercept to finite volume effects visible in the valence chiral
condensate.  We see no evidence for the singular behavior expected from
quenched chiral logarithms.

\end{abstract}

\maketitle

\section{INTRODUCTION}

Over the last few years, analytic calculations using chiral
perturbation theory have indicated pathologies in the quenched
approximation \cite{goltermann,sharpe}.  One example is the lack of
closed quark loops contributing to propagation of the quenched
$\eta^{\prime}$, which leads to a double pole in the quenched
$\eta^{\prime}$ propagator.  This leads to the following
singular behavior for the valence pion mass \cite{sharpe}:
\begin{equation}
  m_{\pi}^2 =  (2 \mu m)^{\frac{1}{1 + \delta}}
    \left( \Lambda^2 \right)^{\frac{\delta}{1 + \delta}}
    \label{eq:mpi_sing}
\end{equation}

For lattice calculations, there is some uncertainty about the mass $m$
to be used in the above.  Due to the flavor symmetry breaking of
staggered fermions at finite lattice spacing, the pions are not
degenerate in mass in current simulations and the quenched
$\eta^{\prime}$ may be closer in mass to the heavier pions than the
lighter ones.  Interpreting the $m$ above as the bare quark mass
assumes the $\eta^{\prime}$ is closer in mass to the lightest,
Goldstone boson, pion \cite{sharpe1}.

Another consequence of the chiral perturbation theory calculations
is the absence of quenched chiral logarithms in decay constants,
in particular in $f_{\pi}$.  Because there is a lattice version
of the Gell-Mann--Oakes--Renner (GMO) relation, this means that
quenched chiral logarithms in the pion mass must be compensated
by quenched chiral logarithms in the quenched chiral condensate.

Recently, there have been reports of simulation data consistent with
the presence of quenched chiral logarithms, particularly in $m_{\pi}$
\cite{kim,gupta}.
To assess whether the data clearly reveals quenched chiral logarithms,
we have investigated three points:
\begin{enumerate}
\item What effects do dynamical fermions have on the data?
  Does the signal
  persist and is it suppressed by the dynamical fermion mass?
\item Are quenched chiral logarithms evident in the
  valence chiral condensate?
\item Have all finite volume effects been clearly controlled?
\end{enumerate}
This note will detail some of our results from the pursuit of
these questions.

\section{RELATING $m_{\pi}^2$ and $\langle \bar{\zeta}\zeta \rangle$ }

Since we will be referring to both quenched and dynamical simulations,
we will introduce a species of staggered fermion, denoted by $\zeta$,
which does not appear in the action.  This valence fermion of mass
$m_{\zeta}$ will be used to construct the valence pion propagators and
the valence chiral condensate.  The mass of the fermions entering the
determinant will be denoted by $m$.

Consider an operator $\pi(x)$ which creates a pion.
We can define a function $C(m_{\zeta})$ by
\begin{equation}
  C(m_{\zeta}) = m_{\pi}^2(m_{\zeta}) \, \int \, d^4x \,
  \langle \pi(x) \pi(0) \rangle \label{eq:c_def}
\end{equation}
The angle brackets can represent averaging over any ensemble of gauge
fields, quenched or dynamical.  Provided the volume is large enough and
$m_{\pi}$ is much less than the mass of any other state with the same
quantum numbers, $m_{\pi}$ in this relation is the mass determining the
decay of the pion propagator.

We now rewrite equation (\ref{eq:c_def}) using the eigenvalues,
$\lambda$, and the density of
eigenvalues per unit volume in a background
configuration $U$, $\rho(\lambda, U)$, of the staggered
fermion Dirac operator, $D(U)$.  We choose our conventions such
that $ D \, \zeta_{\lambda}(x) = i \, \lambda \, \zeta_{\lambda}(x) $
and $\int \, d \lambda \, \rho(\lambda, U) = 1$.  Using the U(1)
symmetry of staggered fermions, we find that the eigenvalues come in
pairs, $\pm \lambda$, and
\begin{equation}
  \int \; d^4x \; \langle \pi(x) \pi(0) \rangle
  =
  2 \int_0^{\infty} \; d\lambda \; \frac{ \bar{\rho}(\lambda, \beta, m)}
  {\lambda^2 + m_{\zeta}^2}
\end{equation}
where $ \bar{\rho}(\lambda, \beta, m) $ is the ensemble average
of $\rho(\lambda, U)$.

We can also write the valence chiral condensate in terms
of the eigenvalue density as
\begin{equation}
 \langle \bar{\zeta}\zeta(m_{\zeta}) \rangle
 =
 2 m_{\zeta}
 \int_0^{\infty} \; d\lambda \; \frac{ \bar{\rho}(\lambda, \beta, m)}
  {\lambda^2 + m_{\zeta}^2} \label{eq:zcc_eigen}
\end{equation}
We therefore have the relation
\begin{equation}
  C(m_{\zeta})
  =
  \frac { m_{\pi}^2(m_{\zeta}) \; \langle \bar{\zeta}\zeta
    (m_{\zeta}) \rangle}
  { m_{\zeta}} \label{eq:gmo}
\end{equation}
If $C(m_{\zeta})$ is known from either simulations or analytic
arguments, this equation connects $m_{\pi}^2$ and
$\langle \bar{\zeta}\zeta(m_{\zeta}) \rangle$ including
any effects of finite volume or quenched chiral logarithms. 

We now focus on equation (\ref{eq:zcc_eigen}) and make a simple model
for the effects of finite volume.  In the continuum, chiral symmetry
breaking is associated with a non-zero value for $\bar{\rho}(0)$, since
$\langle \bar{\zeta}\zeta(0) \rangle = 2 \pi \bar{\rho}(0)$.  The
major effect of finite volume can be postulated to be a minimum
eigenvalue, $\lambda_{\rm min}$, below which the average density
of eigenvalues vanishes.  In particular, consider $\bar{\rho}(\lambda)$
of the form
\begin{equation}
 \bar{\rho}(\lambda) =  \left\{ \begin{array}{ll}
   0		&  \lambda < \lambda_{\rm \min} \\
   \rho_0	&  \lambda_{\rm min} < \lambda < \lambda_0 \\
   \int_{\lambda_0}^{\infty} \; \rho(\lambda) \; d \lambda = 
   \alpha & \lambda_0 < \lambda
   \end{array}
   \right.
\end{equation}
where 
$\alpha$ diverges in the continuum limit due to the quadratic
divergences in $\langle \bar{\zeta}\zeta(m_{\zeta}) \rangle$.
Integrating equation (\ref{eq:zcc_eigen}) for
our model gives,
for $ \lambda_{\rm min} < m_{\zeta} < \lambda_0$,
\begin{equation}
   \langle \bar{\zeta}\zeta(m_{\zeta}) \rangle =
    \rho_0 \left[ \frac{\pi}{2} - \frac{\lambda_{\rm min}}{m_{\zeta}}
      + m_{\zeta}( \alpha - \frac{1}{\lambda_0} )  \right]
  \label{eq:chi_anly}
\end{equation}
plus terms of $O(m_{\zeta}^2)$ and $O(1/m_{\zeta}^2)$.
The model shows that $1/m_{\zeta}$ dependence in
$\langle \bar{\zeta}\zeta(m_{\zeta}) \rangle$ can be
easily associated with finite volume effects.

\pagenumbering{arabic}
\addtocounter{page}{1}

\section{RESULTS FOR $m_{\pi}^2$ and
  $\langle \bar{\zeta}\zeta(m_{\zeta}) \rangle$  }

\begin{figure}[hbt]
\epsfxsize=\hsize
\epsfbox{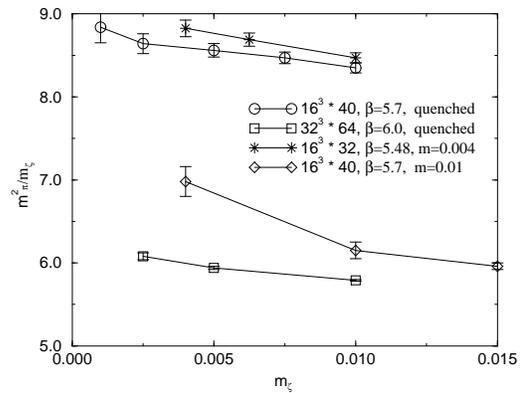}
\caption[fig:mpisq_div_mq]{$m_{\pi}^2/m_{\zeta}$ versus $m_{\zeta}$.}
\label{fig:mpisq_div_mq_vs_mq}
\end{figure}

\begin{figure}
\epsfxsize=\hsize
\epsfbox{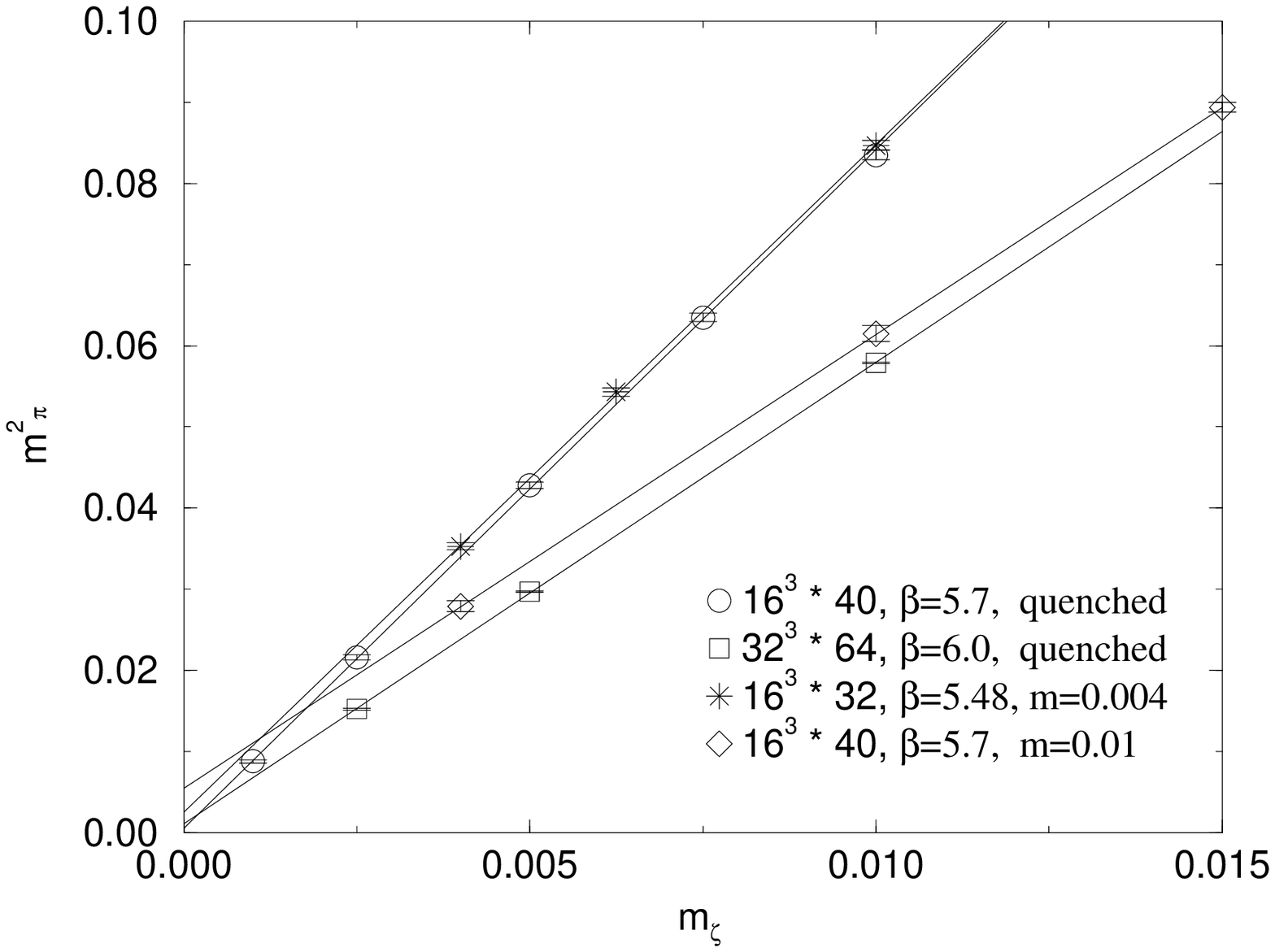}
\caption[fig:mpisq_vs_mq]{$m_{\pi}^2$ versus $m_{\zeta}$.}
\label{fig:mpisq_vs_mq}
\end{figure}

We will be comparing the results from four staggered simulations, two
quenched and two unquenched.  Figure \ref{fig:mpisq_div_mq_vs_mq} shows
$m_{\pi}^2$ versus $m_{\zeta}$ for these four simulations as well as
the simulation parameters.  The $32^3 \times 64$ quenched simulation is
the work of Kim and Sinclair \cite{kim}; the rest were done at
Columbia.  The rise for small $m_{\zeta}$ is the type of behavior
expected from equation (\ref{eq:mpi_sing}).  Unfortunately, this kind of
rise has been evident for some time in simulations and has been widely
attributed to finite volume effects.  In Figure \ref{fig:mpisq_vs_mq}
we replot the data, this time without dividing by $m_{\zeta}$.  The
data is well fit by a straight line, with a non-zero intercept.

The correlated data column of Table \ref{tab:mpisq_fit} gives a list of
the fit parameters we have obtained for fits to $m_{\pi}^2$, without
the covariance matrix, of the form
\begin{equation}
  m_{\pi}^2 = c_0 + c_1 m_{\zeta} \label{eq:mpisq_fit}
\end{equation}
Of course, one should use the full covariance matrix while fitting.  We
have tried single-elimination jackknife fits which simultaneously fit
$m_{\pi}$ for all values of $m_{\zeta}$ and an appropriate range of
times, but we have found such fits to be quite unstable given the large
dimensionality of the covariance matrix.  Instead we have decorrelated
our data.  In particular, we have broken our simulations into three or
four smaller, equal length bins, determined $m_{\pi}$ for the lightest
$m_{\zeta}$ from the first bin, for the next lightest from the second
bin, etc.\ and then done an uncorrelated fit of the form given in
equation (\ref{eq:mpisq_fit}).  The results are given in the column
labeled uncorrelated data in Table \ref{tab:mpisq_fit}.

\begin{table*}[hbt]
\setlength{\tabcolsep}{1.0pc}
\newlength{\digitwidth} \settowidth{\digitwidth}{\rm 0}
\catcode`?=\active \def?{\kern\digitwidth}
\caption{Parameters for linear fits of $m_{\pi}^2$ versus $m_{\zeta}$.}
\label{tab:mpisq_fit}
\begin{tabular*}{\textwidth}{ll|lll|lll}
\hline &
& \multicolumn{3}{c|}{Correlated data} 
& \multicolumn{3}{c}{Uncorrelated data} \\
\multicolumn{1}{c}{$\beta$}  &
\multicolumn{1}{c|}{$ma$}  &
  \multicolumn{1}{c}{ $c_0 \times 10^{-3}$ }	&
  \multicolumn{1}{c}{ $c_1$ }			&
  \multicolumn{1}{c|}{ $\chi^2$/dof }		&
  \multicolumn{1}{c}{ $c_0 \times 10^{-3}$ }	&
  \multicolumn{1}{c}{ $c_1$ }			&
  \multicolumn{1}{c}{ $\chi^2$/dof }		\\ \hline
5.7 & $\infty$ & 0.46(21) & 8.32(5) & 4.5/3 & 0.92(32) & 8.24(7) & 4.2/2
  \\
6.0 & $\infty$	& 1.1(1) & 5.69(2) & 4.1/1 & & & \\ \hline
5.48 & 0.004 & 2.7(8) & 8.27(12) & 0.7/1	& 4.2(8) & 8.01(12) & 1.3/1
  \\
5.7 & 0.01 & 5.5(10)  & 5.59(12) & 0.001/2 & 4.2(14) & 5.69(14) &
  0.3/1 \\ \hline
\end{tabular*}
\end{table*}

\begin{figure}
\epsfxsize=\hsize
\epsfbox{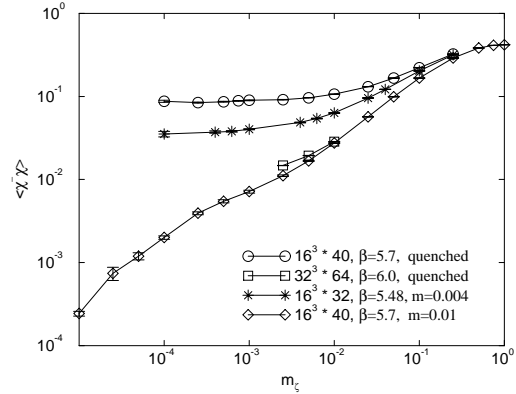}
\caption[fig:chi_zoom]{
$\langle \bar{\zeta}\zeta(m_{\zeta}) \rangle$
versus
$m_{\zeta}$.  The lines on the graph just connect the points.
The $32^3 \times 64$ data has been scaled to the Columbia
normalization.}
\label{fig:chi_zoom}
\end{figure}

We now turn our attention to $\langle \bar{\zeta}\zeta(m_{\zeta})
\rangle$.  Figure \ref{fig:chi_zoom} is a plot of the valence chiral
condensate for the four simulations we are considering.  Notice the
log-log scale and the many orders of magnitude of $m_{\zeta}$ spanned
by the Columbia simulations.

\begin{figure}
\epsfxsize=\hsize
\epsfbox{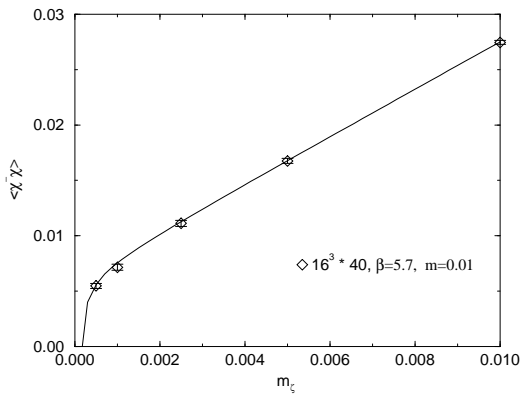}
\caption[fig:chi_m0_01b5_7]
{$\langle \bar{\zeta}\zeta(m_{\zeta}) \rangle$ versus
$m_{\zeta}$ for the $\beta=5.7$, $ma=0.01$ simulation.}
\label{fig:chi_m0_01b5_7}
\end{figure}

Figure \ref{fig:chi_m0_01b5_7} shows $\langle \bar{\zeta}\zeta
(m_{\zeta}) \rangle$  for the $\beta = 5.7$, $ma=0.01$
simulation, on a linear plot.
The line on the graph is a fit of the form
\begin{equation}
  \langle \bar{\zeta}\zeta(m_{\zeta}) \rangle = 
   a_0 + a_1 m_{\zeta} - a_{-1} / m_{\zeta}, \label{eq:zcc_parm}
\end{equation}
motivated by our simple model and equation (\ref{eq:chi_anly}).  For the
correlated fit shown, we find $a_0 = 0.0063(3)$, $a_1 = 2.13(1)$,
$a_{-1} = 8.9(5) \times 10^{-7}$ and $\chi^2/{\rm dof} = 39(16)/2$.
The value of $\chi^2$ is quite large, but the jackknifed error on it is
also large.  This generally indicates a poorly resolved small
eigenvalue in the covariance matrix.  An uncorrelated fit to the data
gives $a_0=0.0061(3)$, $a_1=2.15(2)$, $a_{-1}=-8.8(8) \times 10^{-7}$
and $\chi^2/{\rm dof} = 0.8/2$.  Thus we have clear evidence for finite
volume effects in $\langle \bar{\zeta}\zeta(m_{\zeta}) \rangle$ and
they are in agreement with the expectations of our simple model.

\section{INTERCEPT FOR $m_{\pi}^2$}

We can now use equation (\ref{eq:gmo}) to give us an
understanding of the origin of the non-zero intercept, $c_0$,
found in fitting $m_{\pi}^2$.  Assume that $C(m_{\zeta})$
is a non-singular function in the range $\lambda_{\rm min} < m_{\zeta}
< \lambda_0$.  Then, using the parameterization of
equation (\ref{eq:zcc_parm}) gives
\begin{equation}
  m_{\pi}^2 = - \frac{C(0) a_{-1}}{a_0^2} +
	\left[ \frac{C(0)}{a_0} - \frac{C^{\prime}(0)a_{-1}}
	{a_0^2} \right] m_{\zeta}
\end{equation}
plus terms of $O(m_{\zeta}^2)$ and $O(1/m_{\zeta}^2)$.
This expansion is only valid in the range where $m_{\zeta} >
\lambda_{\rm min}$, which is where most of our data lies.
The non-zero intercept for $m_{\pi}^2$ is easily tied to
the existence of a smallest eigenvalue, $\lambda_{\rm min}$
in the average eigenvalue density $\bar{\rho}$.

One can check that our assumptions about $C(m_{\zeta})$ are
valid in this region by computing the right-hand side of
equation (\ref{eq:gmo}).  We have done this and find
that $C(m_{\zeta})$ shows no singular behavior, so the
power series expansion of the preceeding paragraph appears
reasonable.

Thus we have seen how finite volume effects can account for
a rise in $m_{\pi}^2/m_{\zeta}$ at $\beta=5.7$ and $ma=0.01$.  The
other $16^3 \times N_t$ runs are consistent with the same finite volume
effect, even though they are at much stronger coupling.  This is
reassuring, since the rise in $m_{\pi}^2/m_{\zeta}$ is
apparent for both the quenched and dynamical simulations.
For the $32^3 \times 64$ simulation,
Kim and Sinclair have measured the same value for $m_{\pi}$ on both
$24^3$ and $32^3$ volumes.  Thus they interpret their rise in
$m_{\pi}^2/m_{\zeta}$ as due to a quenched chiral log.  However, their
measurements of $\langle \bar{\zeta}\zeta(m_{\zeta}) \rangle$ do not
show the logarithmic behavior expected, if the rise in
$m_{\pi}^2/m_{\zeta}$ is really a quenched chiral log.  Also,
their data is reasonably well represented by a fit of the
for in equation (\ref{eq:mpisq_fit}).

A possible explanation can come from considering the GMO relation
for infinite volume.  We then have
\begin{equation}
 \frac{m_{\pi}^2}{m_{\zeta}} = \frac{C(0) + C^{\prime}(0) m_{\zeta} }
	{ a_0 + a_1 m_{\zeta} }, \label{eq:gmo_infty}
\end{equation}
if $C(m_{\zeta})$ is a smooth function of $m_{\zeta}$.  Depending on
the relative sizes of $a_1/a_0$ and $C^{\prime}(0)/C(0)$, one could see
a rise in $m_{\pi}^2/m_{\zeta}$ without finite volume effects or chiral
logarithms, until $m_{\zeta}$ is very small.  The relative sizes of
$a_1/a_0$ and $C^{\prime}(0)/C(0)$ are generally $\beta$ dependent, so
that even at infinte volume $m_{\pi}^2/m_{\zeta}$ could have a
different form as a function of $\beta$.  In addition, if one includes
the $a_{-1}$ term in equation (\ref{eq:gmo_infty}), there are choices
for the parameters which can give a decrease in $m_{\pi}^2/m_{\zeta}$
for moderate values of $m_{\zeta}$, and then a rise for small
$m_{\zeta}$.  Such behavior has been reported by S. Gottlieb \cite{sg}
at this conference.

We would like to thank S. Kim and D. Sinclair for a preliminary
copy of their paper.  We also thank Pavlos Vranas and Roy Luo
for helpful discussions.

\end{document}